\documentclass[aps,12pt,tightenlines,amsmath,amssymb]{revtex4}
\usepackage{graphicx}
\usepackage{dcolumn}
\usepackage{bm}

\begin{document}

\title{AXION-LIKE PARTICLES AND e-ASTROGAM}

\author{Marco Roncadelli$^{a}$, Giorgio Galanti$^{a}$, Alessandro De Angelis$^b$}
\affiliation{$(a)$ INFN, Sezione di Pavia, Via A. Bassi 6, I -- 27100 Pavia, Italy, and INAF\\
$(b)$ INFN and INAF Padova; Univerity of Udine; LIP Lisboa}

\begin{abstract}

\noindent {\bf{Abstract.}} e-ASTROGAM \cite{ea} is a space mission with unprecedented sensitivity to photons in the MeV range, proposed within the ESA M5 call. In this note we describe some 
measurements sensitive to axion-like particles, for which performance in the MeV/GeV range is of primary importance, and e-ASTROGAM could be the key for discovery.


\end{abstract}


\maketitle

\begin{figure}[h]
\centering
\includegraphics[width=0.27\textwidth]{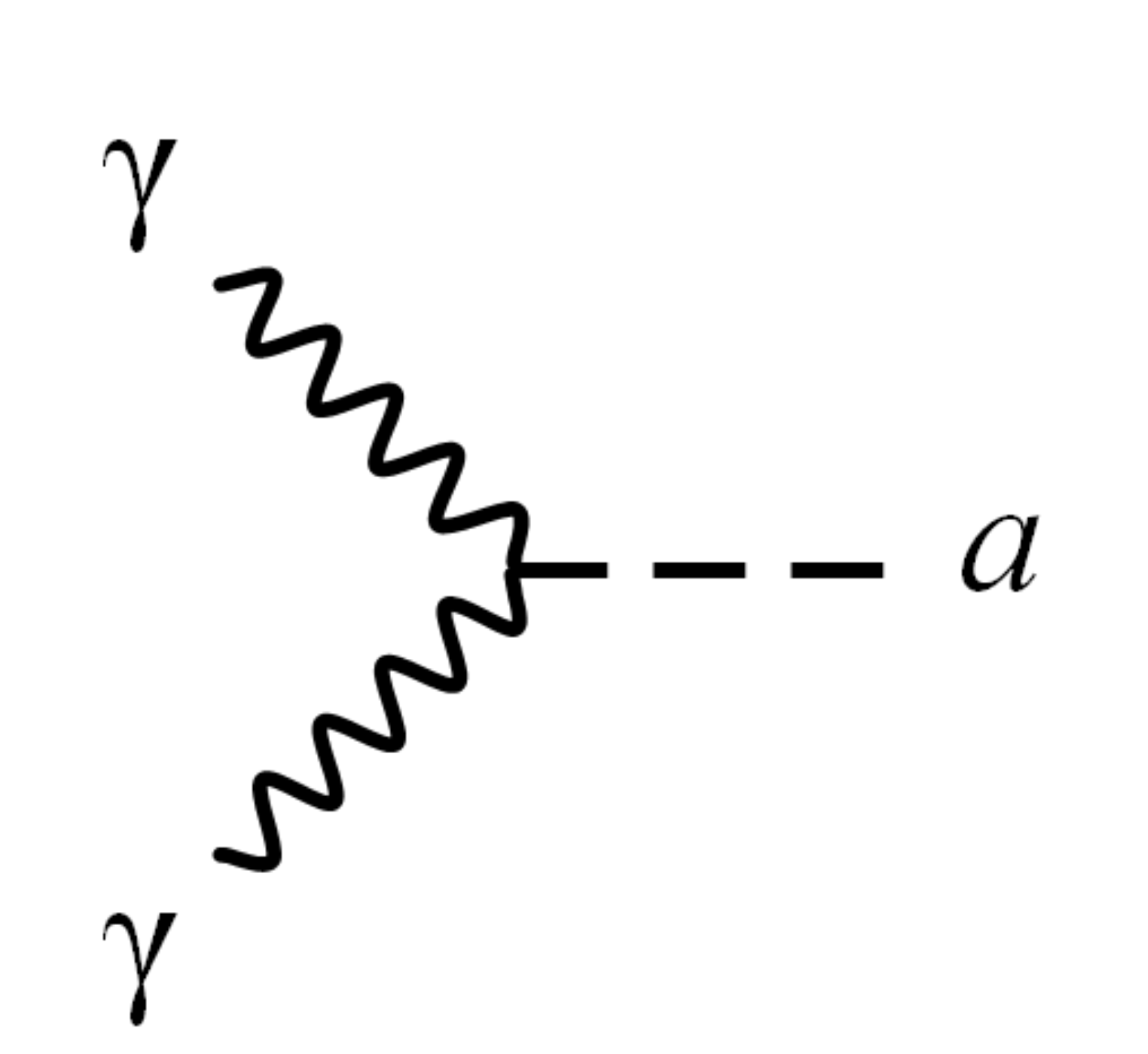}
\caption{\label{immagine3(2)} Photon-photon-ALP vertex.}
\end{figure}

Axion-like particles (ALPs) are neutral and very light pseudo-scalar particles $a$ (see \cite{alp1} and references therein for a review of the properties of axions and ALPs). They are a generic prediction 
of many extensions of the Standard Model, especially of those based on superstrings (in a broad sense). They are similar to the axion apart from two features. First, ALPs couple almost only to two photons through $g_{a \gamma} \, a \, {\bf B} \cdot {\bf E}$ (very small couplings to fermions are allowed but here they are discarded because they do not give rise to any interesting effect). Second, the two-photon coupling $g_{a \gamma}$ is totally unrelated to the ALP mass $m$. Hence, ALPs are described by the Lagrangian 
\begin{equation}
\label{t1}
{\cal L}_{\rm ALP} = \frac{1}{2} \, \partial^{\mu} a \, \partial_{\mu} a - \, \frac{1}{2} \, m^2 \, a^2 + g_{a \gamma} \, a \, {\bf E} \cdot {\bf B}~,
\end{equation}
where ${\bf E}$ and ${\bf B}$ denote the electric and magnetic components of the field strength $F^{\mu \nu}$. So, for ALPs the only new thing with respect to the Standard Model is shown in Figure~\ref{immagine3(2)}.

\begin{figure}[h]
\centering
\includegraphics[width=0.30\textwidth]{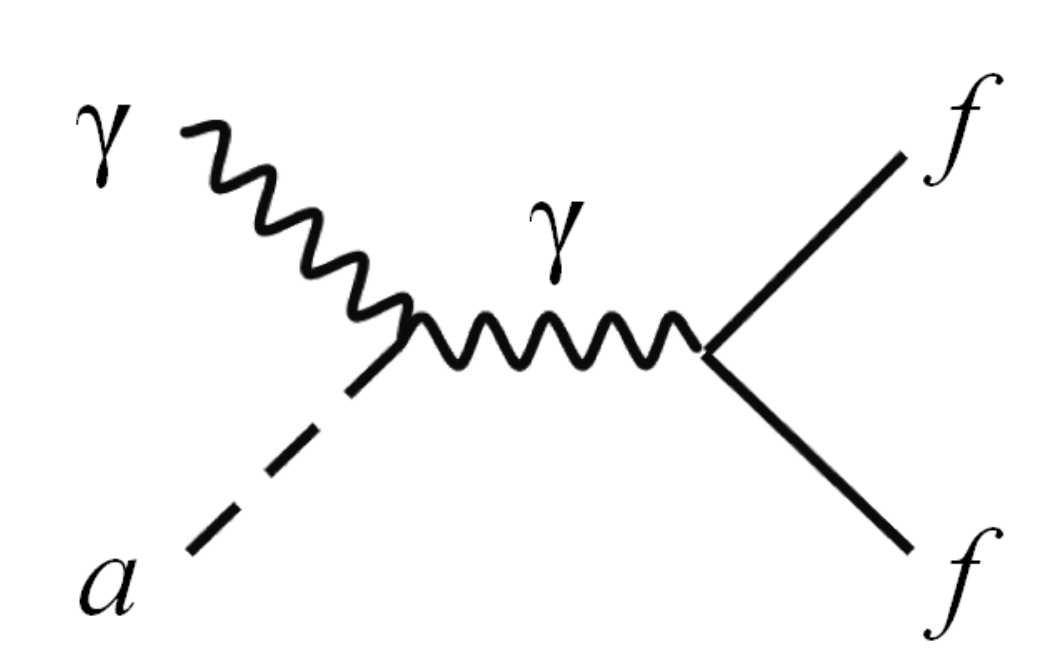}
\caption{\label{fey5} This Feynman diagram represents the $a \gamma \to f {\overline f}$ scattering in the $s$-channel and the $a f \to \gamma f$ scattering in the $u$-channel.}
\end{figure}

\begin{figure}[h]
\centering
\includegraphics[width=.30\textwidth]{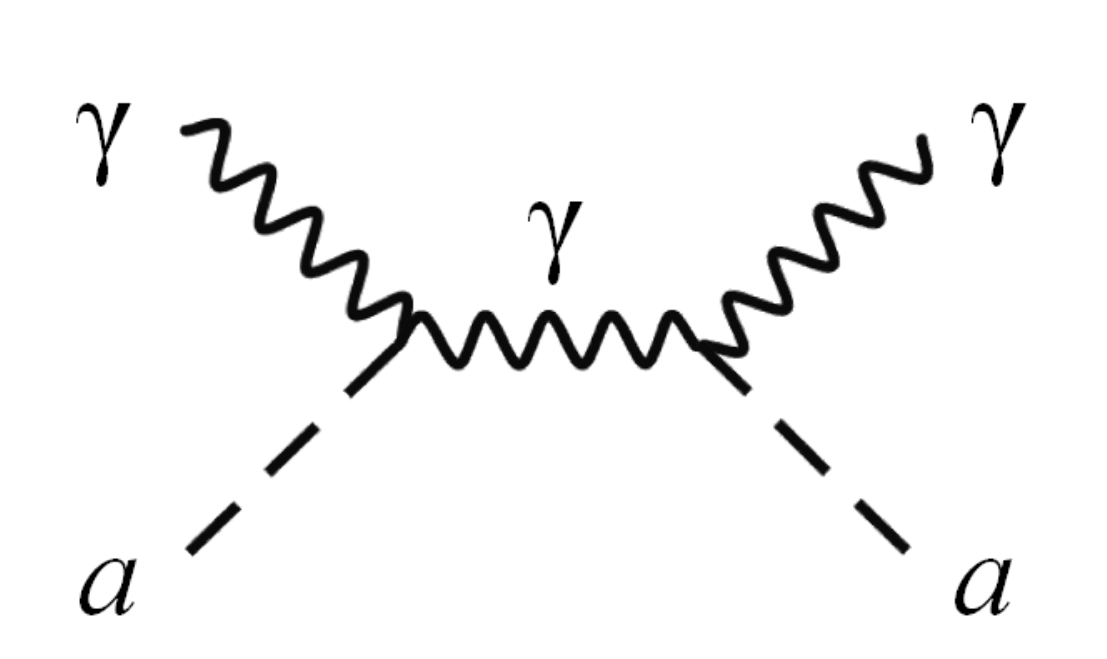}
\caption{\label{ALPph} Feynman diagram for the $a \gamma \to a \gamma$ scattering.}
\end{figure}

In spite of the fact that ALPs couple to two-photons, their interaction with both matter and radiation is negligibly small. To see 
this, consider the Feynman diagram in Figure~\ref{fey5}, where $f$ is a generic fermion.~ 
Since the cross-section is $\sigma \sim \alpha \, g^2_{a \gamma}$ we indeed get $\sigma < 10^{- 50} \, {\rm cm}^2$. Moreover, let us address $a \gamma \to a \gamma$ scattering depicted in Figure~\ref{ALPph}.~ 
Then the cross-section is $\sigma \sim s \, g^4_{a \gamma}$, and so we find 
\begin{equation}
\label{t1q}
\sigma < 7 \cdot 10^{- 69} \left(\frac{s}{{\rm GeV}^2} \right) {\rm cm}^2~.
\end{equation}

\section{Conversion of photons into axions through cosmic propagation}

\begin{figure}[h]
\centering
\includegraphics[width=.30\textwidth]{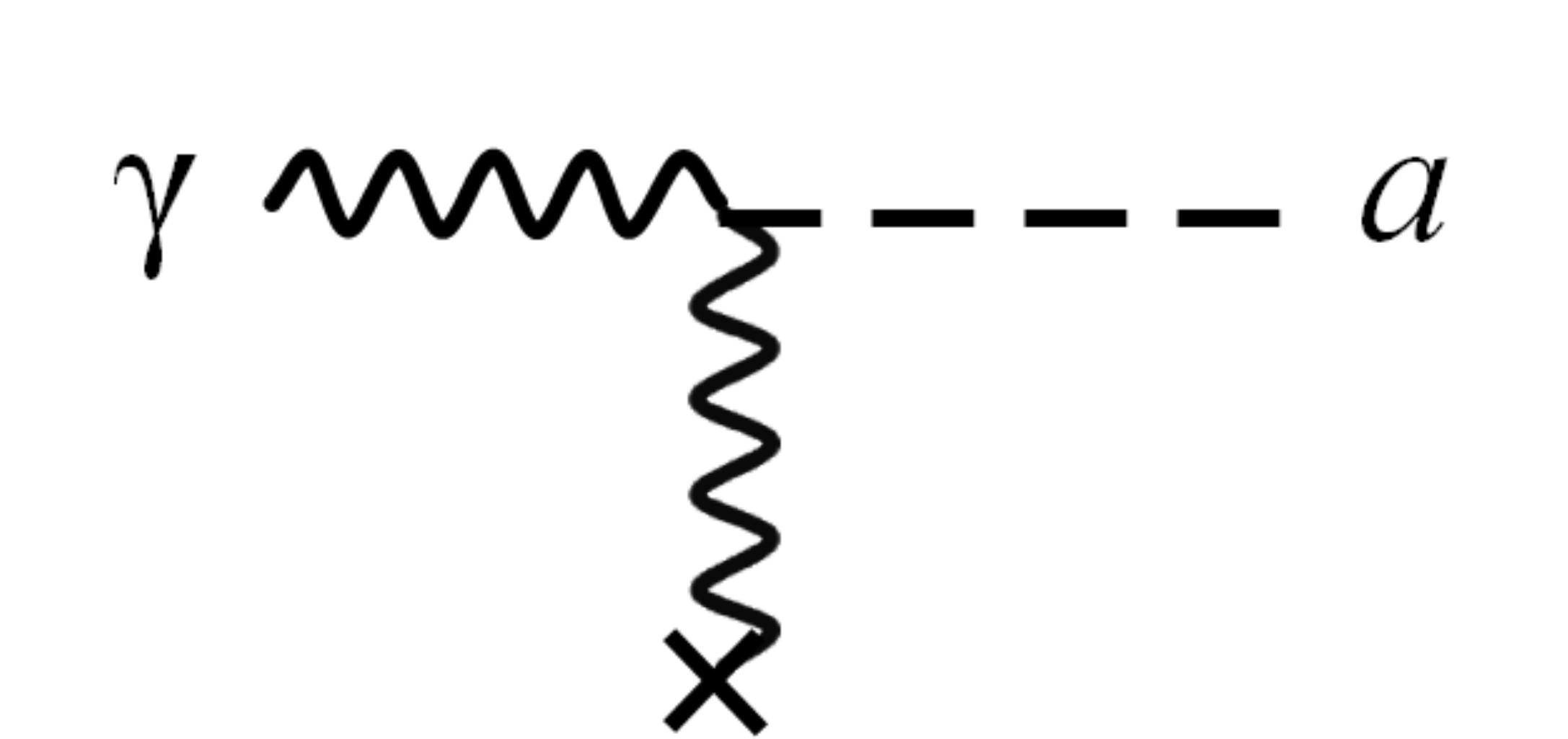}
\caption{\label{immagine4(2)} $\gamma \to a$ conversion in the external magnetic field ${\bf B}$ (in stars the 
r\^ole of ${\bf E}$ and ${\bf B}$ is interchanged).}
\end{figure}

We will henceforth consider a monochromatic photon beam and assume that an external magnetic field ${\bf B}$ is present (in stars the r\^ole of ${\bf E}$ and ${\bf B}$ is interchanged, but we shall not be interested in this situation). Hence in $g_{a \gamma} \, a \, {\bf E} \cdot {\bf B}$ the term ${\bf E}$ is the electric field of a beam photon while ${\bf B}$ is the external magnetic field. So the mass matrix in the photon-ALP sector is non-diagonal, which implies that $\gamma \to a$ conversions occur, like the one shown in Figure~\ref{immagine4(2)}.

\begin{figure}[h]
\centering
\includegraphics[width=0.6\textwidth]{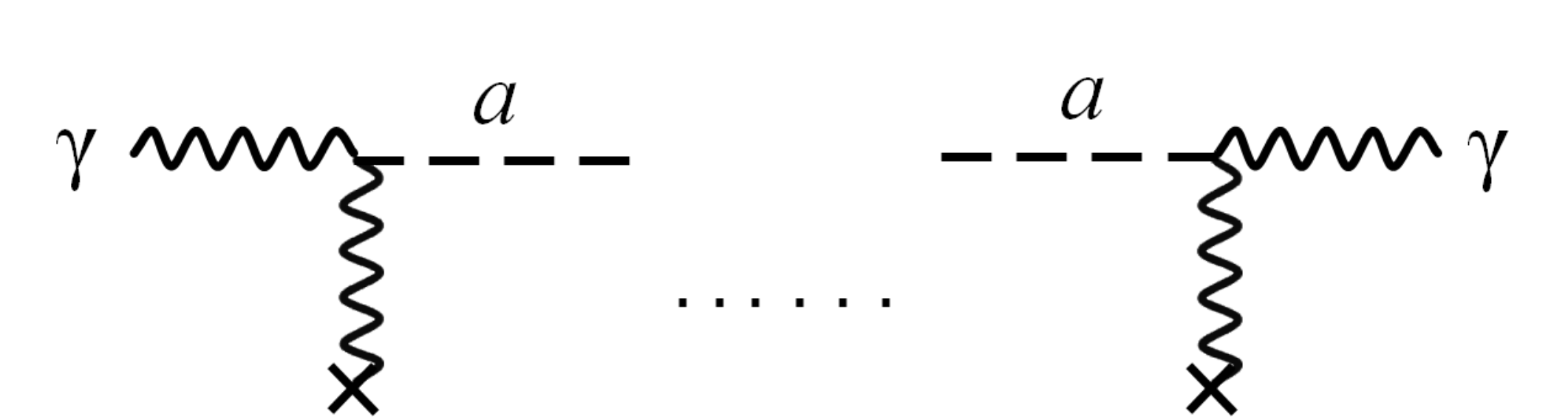}
\caption{\label{fey1} Schematic view of a photon-ALP oscillation in the external magnetic field ${\bf B}$.}
\end{figure}

\noindent Of course, also the inverse process $a \to \gamma$ takes place, an so as the beam propagates several $\gamma \to a$ and $a \to \gamma$ transitions occur, like in Figure~\ref{fey1}, where $a$ is real. Thus, in the presence of an external ${\bf B}$ field photon-ALP {\it oscillations} $\gamma \leftrightarrow a$ can occur much in the same way as it happens for massive neutrinos of different flavor, apart from the need of an external field in order to compensate for the spin mismatch.      

\begin{figure}[h]
\centering
\includegraphics[width=0.6\textwidth]{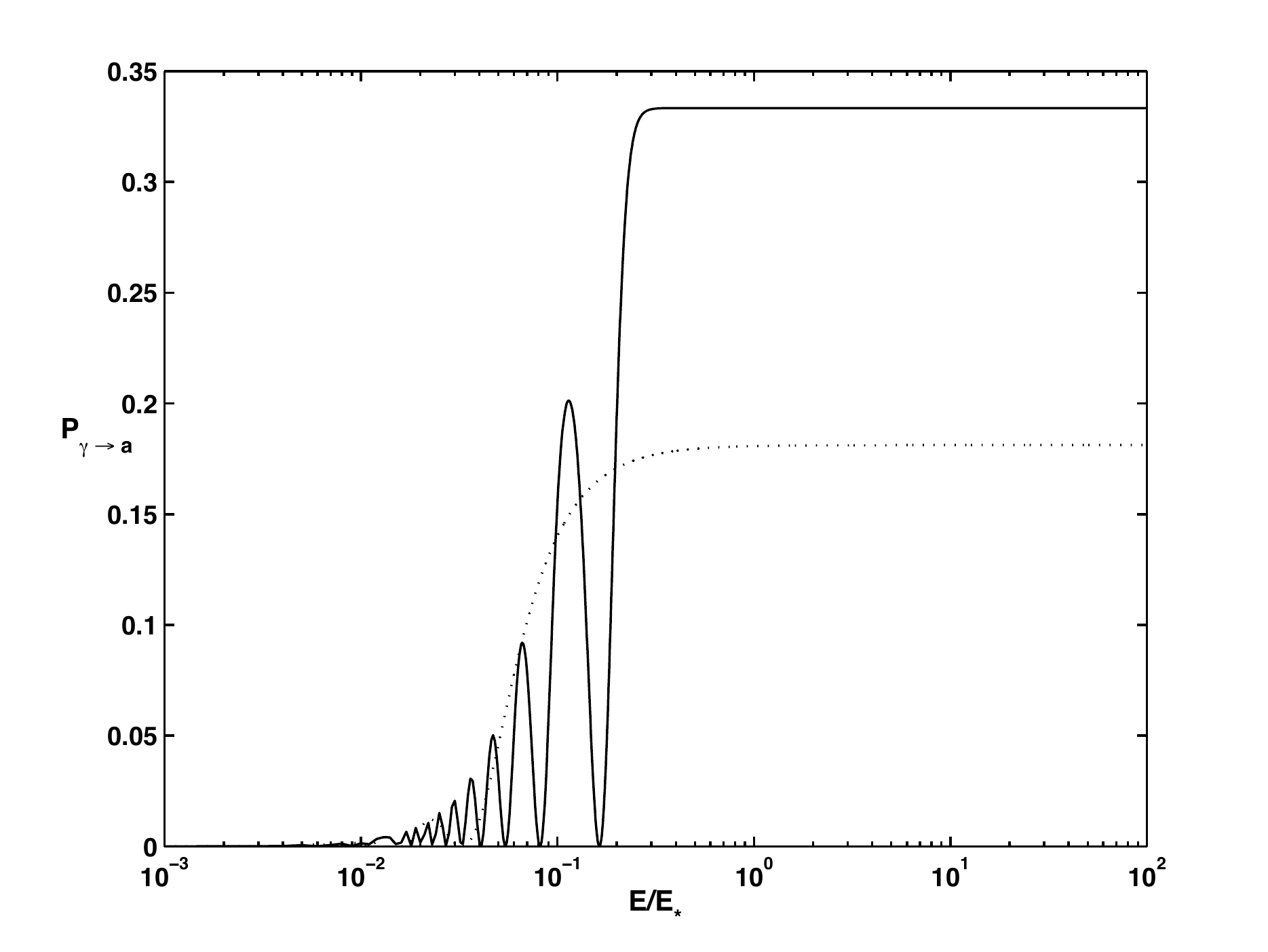}
\caption{\label{marco} Oscillatory behavior around $E_*$ for $g_{a \gamma} = 0.33 \cdot 10^{- 10} \, {\rm GeV}^{- 1}$, 
$0.5 \, {\rm nG}$ and $N = 200$ magnetic domains.}
\end{figure}

The key-point to detect ALPs with e-ASTROGAM is as follows. Suppose that a distant source emits a  
$\gamma/a$ beam of energy $E$ in the range $0.3 \, {\rm MeV} - 3 \, {\rm GeV}$ which propagates along the $y$ direction reaching us. In the approximation $E \gg m$ -- which is presently valid -- the beam propagation equation becomes a 
Schr\"odinger-like equation in $y$, hence the beam is {\it formally} described as a {\it 3-level quantum system}. Consider now the simplest possible case, where no photon absorption takes place and $\bf B$ is homogeneous. 
Taking ${\bf B}$ along the $z$-axis, we have (see for example \cite{noi} for a review of the assumptions and the details of the calculations)
\begin{equation} 
P_{\gamma \to a} (E; 0, y) = \left(\frac{g_{a \gamma} \, B}{\Delta_{\rm osc}} \right)^2 \, {\rm sin}^2 \left( \frac{\Delta_{\rm osc} \, y}{2} \right)~, \ \ \ \  {\Delta}_{\rm osc} \equiv \left[\left(\frac{m^2 - {\omega}_{\rm pl}^2}{2 E}  \right)^2 + \bigl(g_{a \gamma} \, B \bigr)^2 \right]^{1/2}~,
\label{a16ghA}
\end{equation}
where ${\omega}_{\rm pl}$ is the plasma frequency of the medium. Defining next $E_* \equiv | m^2 - {\omega}^2_{\rm pl} |/(2 \, g_{a \gamma} \, B)$ it turns out that $P_{\gamma \to a} (E; 0, y) = 0$ for $E \ll E_*$, $P_{\gamma \to a} (E; 0, y)$ rapidly oscillates with $E$ for $E \sim E_*$ -- this is the {\it weak-mixing regime} -- while $P_{\gamma \to a} (E; 0, y)$ is maximal and independent of $m$ and $E$ for $E \gg E_*$, which is the {\it strong-mixing regime}. Actually, the extragalactic magnetic field ${\bf B}$ is currently modeled as a domain-like structure with $L_{\rm dom} = (1 - 10) \, {\rm Mpc}$, $B = (0.1 - 1) \, {\rm nG}$ in all domains, but the ${\bf B}$ direction changes randomly in any domain: this ${\bf B}$ structure is strongly motivated by galactic outflow models, and it turns out to enhance the oscillatory regime around $E_*$ as shown in Figure~\ref{marco}. 

We stress that -- on top of the above oscillatory behavior -- we also have a slight dimming of the signal \cite{dmr2008}: because of energy conservation, the production of ALP implies a reduced photon flux. Actually, it can be shown that for $N \gg 1$ magnetic domains, the two photon polarization states and the single ALP state undergo equipartition, so that the signal becomes dimmer by a factor of 0.66. In conclusion, the two features showing up in the energy spectrum due to an ALP are an {\it oscillatory behavior} and a {\it dimming of 0.66}. 

Yet, this is not the end of the story, because e-ASTROGAM can perform also polarimetric measurements. This is because the coupling $g_{a \gamma} \, a \, {\bf E} \cdot {\bf B}$ acts as a {\it polarizer}. It is indeed trivial to see that it entails that only the component ${\bf B}_T$ orthogonal to the photon momentum ${\bf k}$ matters, and that photons $\gamma_{\perp}$ with linear polarization orthogonal to the plane defined by ${\bf k}$ and ${\bf B}$ do not mix with $a$, but only photons $\gamma_{\parallel}$ with linear polarization parallel to that plane do \cite{mpz1986}. Specifically, two distinct phenomena come about: {\it birefringence}, namely the change of a linear polarization into an elliptical one with the major axis parallel to the initial polarization, and {\it dichrois}, namely a selective conversion $\gamma \to a$ which causes the ellipse's major axis to be misaligned with respect to the initial polarization. Thus, the measure of the polarization of radiation with known initial polarization provides additional information to discriminate an ALP from other possible effects. Very remarkably, we actually do {\it not need} to know the initial polarization by employing a simple trick. Because when one does not measure the polarization one has to sum over the two final photon polarizations -- while when one does measure it no sum is performed -- the signal has to be {\it twice as large} when the polarization is not measure as compared with the case in which the polarization is measured.
 
Finally, what is the mass range of the ALP that can be probed by e-ASTROGAM? It follows from the previous considerations that the answer is given by $E \sim E_*$. Neglecting ${\omega}_{\rm pl}$ and recalling the second of Eqs. (\ref{a16ghA}), such 
a condition becomes 
\begin{equation} 
E \sim \frac{m^2}{2 g_{a \gamma} \, B}~, 
\label{a16ghAB}
\end{equation}
Owing on the energy range of e-ASTROGAM, from Eq. (\ref{a16ghAB}) we have
\begin{equation} 
0.3 \, {\rm MeV} < \frac{m^2}{2 g_{a \gamma} \, B} < 3 \, {\rm GeV}~. 
\label{a16ghAC}
\end{equation}
Taking e.g. $g_{a \gamma} = 0.33 \cdot 10^{- 10} \, {\rm GeV}^{- 1}$ and $B = 0.5 \, {\rm nG}$ the answer is
\begin{equation} 
1.7 \cdot 10^{- 13} \, {\rm eV} < m <  1.7 \cdot 10^{- 11} \, {\rm eV}~. 
\label{a16ghAD}
\end{equation}
More generally, by employing the parametrizations $g_{a \gamma} = \alpha \, 10^{- 10} \, {\rm GeV}^{- 1}$ and $B = \beta \, 
{\rm nG}$, we get (independent of $N$)
\begin{equation} 
1.1 \cdot 10^{- 12} \, \alpha \, \beta \, {\rm eV} < m <  1.1 \cdot 10^{- 10} \, \alpha \, \beta \, {\rm eV}~. 
\label{a16ghAE}
\end{equation}

\section{Axion-like particles from  from Supernovae}

ALPs are produced in the core of  stars (like the Sun) through the Primakoff process in the Coulomb field ${\bf E}$ of ionized matter, illustrated in Figure~\ref{immagine4(2)}. The CAST experiment at CERN was looking at the Sun and found nothing, thereby deriving $g_{a \gamma} < 0.88 \cdot 10^{- 10} \, {\rm GeV}^{- 1}$. Recent analysis of globular clusters gives $g_{a \gamma} < 0.66 \cdot 10^{- 10} \, {\rm GeV}^{- 1}$ \cite{pdg2016}. 
They can also be produced in at the centre of core-collapse (Type II) supernovae soon after the bounce (when also the neutrino burst is produced) by Primakoff effect
and reconverted to photons of the same energy, during their travel or in the Milky Way. The arrival time of these photons would
be the same as for neutrinos, providing a clear signature.

Integrated over the explosion time, which is of the order of 10\,s, the Authors of  \cite{meyer2017}
 find that the ALP spectrum can be parametrized by a power law with exponential cutoff,
\begin{equation}
\frac{dN_a}{dE} = C \left(\frac{g_{a\gamma}}{10^{-11}{\rm{GeV}}^{-1}}\right)^2
\left(\frac{E}{E_0}\right)^\beta \exp\left( -\frac{(\beta + 1) E}{E_0}\right) \,\, 
\label{eq:time-int-spec}
\end{equation}
with $ C $, $ E_0 $, and $ \beta $ given in Tab.~\ref{tab:rate_parameters}.
\begin{table}[h]
	\begin{center}
		\begin{tabular}{ l  c  c   c  }
			\hline \hline
			Progenitor mass			& $ C $ [$10^{50}${MeV}$^{-1} $]		&  $ E_0 $ [MeV] 					& $ \beta  $    \\ \hline 
			$10\,M_\odot$	& $ 5.32$  	& $ 94 ~ $  	& 2.12		\\ 
			$18\,M_\odot$	&	$ 9.31 $	& $102 ~ $   	& 2.25  \\ \hline
		\end{tabular}
		\caption{Best fitting values for the parameters in Eq.~\eqref{eq:time-int-spec}.}
		\label{tab:rate_parameters}   
	\end{center}
\end{table}

\begin{figure}[h]
\centering
\includegraphics[width=0.4\textwidth]{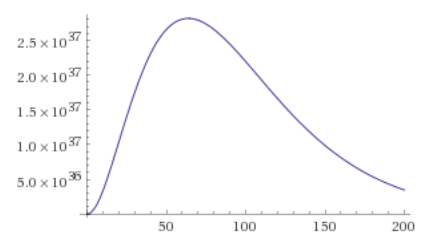}
\includegraphics[width=0.4\textwidth]{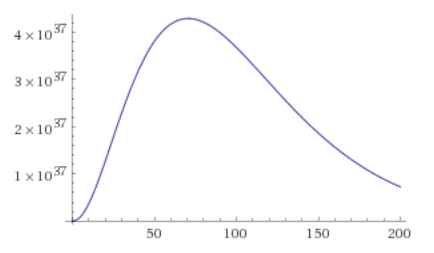}
\caption{\label{fig:alpsn} The differential axion rate from the supernova, $dN_a/dE ({\rm{GeV}}^{-1})$, for a SN of 10 (left) and 18 (right) solar masses. The abscissa is in MeV.}
\end{figure}

The ALP energy spectrum (which corresponds to the photon energy spectrum after reconversion to gamma-rays) is shown in Figure \ref{fig:alpsn}.  The bulk is below $\sim$100 MeV, which clearly shows the potential of e-ASTROGAM for a possible detection.  e-ASTROGAM
would have a sensitivity better than Fermi-LAT and access to much
smaller mass/coupling values than dedicated laboratory experiments.

\begin{thebibliography}{99}
\bibitem{ea} A. De Angelis, V. Tatischeff, M. Tavani et al., ``The e-ASTROGAM mission'', arXiv:1611.02232, to be published in Experimental Astronomy (2017)
\bibitem{alp1} J. Jaeckel and A. Ringwald,  Ann. Rev. Nucl. Part. Sci. {\bf 60}, 405 (2010)
\bibitem{noi} A. De Angelis, G. Galanti, and M. Roncadelli, Phys. Rev.
D {\bf 84}, 105030 (2011); {\em erratum,} Phys. Rev. D {\bf 87}, 109903 (2013)
\bibitem{dmr2008} A. De Angelis, O. Mansutti and M. Roncadelli, Phys. Lett. B {\bf 659}, 847 (2008)
\bibitem{mpz1986} L. Maiani, R. Petronzio and E. Zavattini, Phys. Lett. B {\bf 175}, 359 (1986) 
\bibitem{pdg2016} C. Patrignani et al. (Particle Data Group), Chin. Phys. C {\bf 40}, 100001 (2016)
\bibitem{meyer2017} M. Meyer et al., Phys. Rev. Lett. {\bf 118}, 011103 (2017)
\end{thebibliography}
\end{document}